\documentclass[aps,prx,twocolumn,
               amsmath,amssymb,amsfonts,
               notitlepage]{revtex4-1}
\AtBeginDocument{\usepackage{booktabs}}               

\usepackage{microtype}
\usepackage{wasysym}
\usepackage[varg]{txfonts}
\usepackage{hyperref}
\usepackage{amsmath}
\usepackage{color}
\usepackage{graphicx}
\usepackage[percent]{overpic}
\usepackage{mathrsfs}
\usepackage{bm}
\usepackage{braket}

\newcommand{\tsub}[1]{\textsubscript{#1}}

\newcommand{\del} {\partial}

\newcommand{\h}[1]{{#1}^{\dagger}}
\renewcommand{\k}[1]{{#1}^{}}

\newcommand{\re}{{\rm Re}}

\newcommand{\hc}{{\rm h.c.}}
\newenvironment{subalign}{\subequations\align}{\endalign\endsubequations}

\renewcommand{\vec}[1]{\boldsymbol{#1}}
\newcommand{\mat}[1]{\vec{#1}}
\newcommand{\trp}[1]{{#1}^{\intercal}}
\newcommand{\vhat}[1]{\vec{\hat{#1}}}

\newcommand{\sectitle}[1]{\emph{#1}:}

\newcommand{\abo}[2]{#1\tsub{2}#2\tsub{2}O\tsub{7}}
\newcommand{\eto}{\abo{Er}{Ti}}

\newcommand{\mueV}{\ \mu{\rm eV}}
\newcommand{\meV}{\ {\rm meV}}

\definecolor{cL}{RGB}{59, 83, 140}
\definecolor{cM}{RGB}{33, 145, 141}
\definecolor{cH}{RGB}{95, 202, 98}
\definecolor{cG}{RGB}{204,204,204}

\newcommand{\garnet}{Ca\tsub{3}Fe\tsub{2}Ge\tsub{3}O\tsub{12}}
\hypersetup{colorlinks=true,linkcolor=cL,citecolor=cL,urlcolor=cL} 

\usepackage[nolist,nohyperlinks]{acronym}
\newacro{DM}[DM]{{Dzyaloshinskii-Moriya}}
\newacro{SM}[SM]{Supplemental Material}

\begin{document}

\title{Pseudo-Goldstone gaps and order-by-quantum-disorder in frustrated magnets}
\author{Jeffrey G. Rau}
\author{Paul A. McClarty}
\author{Roderich Moessner}
\affiliation{Max-Planck-Institut f\"ur Physik komplexer Systeme, 01187 Dresden, Germany}
\begin{abstract}
In systems with competing interactions, continuous degeneracies can
appear which are accidental, in that they are not related to any
symmetry of the Hamiltonian. Accordingly, the pseudo-Goldstone modes
associated with these degeneracies are also unprotected. Indeed,
through a process known as ``order-by-quantum-disorder'', quantum zero-point
fluctuations can lift the degeneracy and induce a gap for these
modes. We show that this gap can be exactly computed at leading order
in $1/S$ in spin-wave theory from the mean curvature of the classical and quantum zero-point
energies -- without the need to consider any spin-wave
interactions. We confirm this equivalence through direct calculations
of the spin-wave spectrum to $O(1/S^2)$ in a wide variety of
theoretically and experimentally relevant quantum spin models.
We prove this equivalence through the use
of an exact sum rule that provides the required mixing of different
orders of $1/S$.  Finally, we discuss
some implications for several leading order-by-quantum-disorder candidate
materials, clarifying the expected pseudo-Goldstone gap sizes in
\eto{} and \garnet{}.
\end{abstract}

\date{\today}

\maketitle
Goldstone's theorem~\cite{goldstone1962} connects the spontaneous
breaking of a continuous symmetry to the presence of gapless
excitations -- a foundational result with applications in almost every
branch of physics. Alternatively, gapless excitations can be generated
by \emph{accidental} degeneracies which are not symmetry
enforced~\cite{weinberg1972pseudo}. Just as continuous symmetries
imply the presence of gapless Nambu-Goldstone modes, accidental degeneracies
imply the presence of pseudo-Goldstone modes that are
\emph{nearly} gapless when (inevitably) these degeneracies are weakly
lifted. Such modes have been invoked to explain the appearance of
unexpectedly low-lying excitations many contexts, ranging from from
quantum chromodynamics~\cite{weinberg1995quantum} to high-temperature
superconductors~\cite{zhangso5,fernandes2016low} and quantum
magnets~\cite{villain1980order,henley1989,shender1982antiferromagnetic};
perhaps the most well known example is the mass of the pion, which arises
due to 
broken chiral symmetry~\cite{nambu1,nambu2}.

In lieu of explicit symmetry breaking, accidental degeneracies can
also be lifted by \emph{fluctuations}.  Broadly referred to as
``order-by-disorder''~\cite{villain1980order,henley1989,shender1982antiferromagnetic},
this phenomenon has proven useful in understanding a wide variety of
ordering phenomena in frustrated spin
systems~\cite{chandracoleman1990,moessner1998geom,chandradoucot1988,henley1989,chubukov1992triangular},
where accidental degeneracies are natural.
An example is ``order-by-quantum-disorder''~\cite{shender1982antiferromagnetic,henley1989},
where an accidentally degenerate manifold in the classical limit,
$S\rightarrow \infty$, is lifted by quantum corrections at $O(1/S)$.
Within non-interacting spin-wave theory~\cite{holstein1940lswt}, these
contributions can be viewed as the zero-point energy of the harmonic
spin-waves selecting some subset of the classically degenerate
manifold~\cite{shender1982antiferromagnetic,henley1989}.

Through order-by-quantum-disorder, the pseudo-Goldstone modes
associated with this accidental degeneracy must acquire a gap. Since
in non-interacting spin-wave theory these modes are gapless, to obtain
a finite gap one must include the effects of spin-wave
interactions. While conceptually simple, such calculations have only
been carried out for a few limited cases, mostly for simple isotropic
Heisenberg-like models~\cite{belorizky1980calculation,
shender1982antiferromagnetic,
chubukov1992triangular,harris2001quantum,
yildrim1998frustration,aczel2016fcc,pg-tymo}.  In more complex models with
strong exchange anisotropy, such as in the rare-earth
pyrochlores~\cite{gardner2010rmp} or in Kitaev
magnets~\cite{rau2016spin,winter2017models}, these calculations can be
complicated by the presence of three-magnon interactions that can lead to
spontaneous magnon decay~\cite{zhitomirsky2013rmp}, even for colinear
magnetic ground states.
\begin{figure}[tp]
  \begin{overpic}[width=\columnwidth]{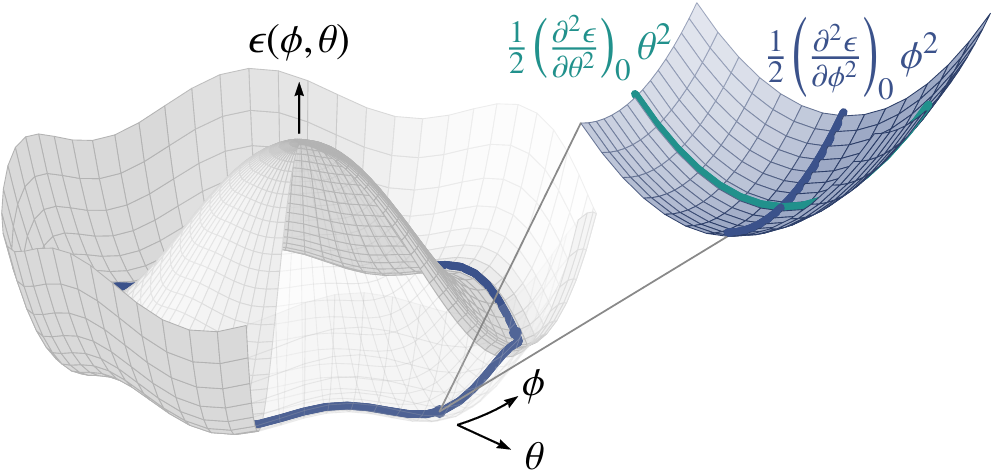}
  \end{overpic}
  \caption{\label{fig:obd}
    Schematic illustration of the semi-classical energy density
$\epsilon(\phi,\theta)$, and the curvatures $\left(\frac{\del^2
\epsilon}{\del \theta^2}\right)_0$ and $\left(\frac{\del^2
\epsilon}{\del \phi^2}\right)_0$ about the semi-classical ground state, which are
related to the pseudo-Goldstone gap, $\Delta$ via
Eq.~(\ref{eq:curvature-formula}). The nearly soft manifold associated
with a type I pseudo-Goldstone mode is shown, with the two principal
curvatures about a global minimum indicated.  }
\end{figure}

In this paper, we offer a significant simplification, showing that the
curvatures of classical and quantum zero-point energy densities
computed at $O(1/S)$, are already sufficient to determine the
pseudo-Goldstone gap exactly to $O(1/S^2)$. Explicitly, if the
classically degenerate manifold is parameterized by $\phi$ with
conjugate direction $\theta$ (see Fig.~\ref{fig:obd}), the
pseudo-Goldstone gap, $\Delta$, is given by~\cite{supp}
\begin{equation}
  \label{eq:curvature-formula}
  \Delta = \frac{1}{S}\sqrt{\left(\frac{\del^2 \epsilon}{\del \theta^2}\right)_0\left(\frac{\del^2 \epsilon}{\del \phi^2}\right)_0 - \left(\frac{\del^2 \epsilon}{\del \theta \del \phi}\right)_0^2},
\end{equation}
where the semi-classical energy density, $\epsilon(\phi,\theta)$, of
the classical ground state at $(\phi,\theta)$ includes the classical
[$O(S^2)$] and quantum zero-point [$O(S)$] contributions.
While used as a heuristic in several
works~\cite{belorizky1980calculation,shender1982antiferromagnetic,jackeli2009mott,
  chaloupka2010kh,savary2012obd}, its equivalence
to the leading result from non-linear spin-wave theory is far from
evident in perturbation theory, as it involves mixing of different
orders in $1/S$. This formula eliminates much of burden of computing
the pseudo-Goldstone gap, requiring only quantities from standard
non-interacting spin-wave theory
a computation considerably more
straightforward to undertake in practice.

In light of this, we revisit a variety of models that exhibit
order-by-quantum-disorder, including square and cubic
Heisenberg-compass models~\cite{belorizky1980calculation},
Heisenberg-Kitaev-$\Gamma$ models~\cite{chaloupka2010kh,rau2014jkg} on
the honeycomb lattice and $J_1-J_2$ models on the square and
triangular
lattices~\cite{henley1989,chandradoucot1988,jolicoeur1990,chubukov1992triangular}.
For each, we compute the gap both explicitly in interacting
spin-wave theory -- often for the first time -- and then again using
the curvature formula [Eq.~(\ref{eq:curvature-formula})], confirming
that they are indeed identical.

Finally, we consider applications; while order-by-quantum-disorder has a long theoretical
history~\cite{tessman,belorizky1980calculation,villain1980order,shender1982antiferromagnetic,henley1989},
there are only a handful of serious potential experimental
candidates~\cite{shender1982antiferromagnetic,brueckel1988dynamical,kim1999obd,harris2001quantum,savary2012obd,zhitomirsky2012obd}.
Two of the best material examples are the cubic Heisenberg
anti-ferromagnet
\garnet{}~\cite{shender1982antiferromagnetic,brueckel1988dynamical},
and the pyrochlore XY anti-ferromagnet
\eto{}~\cite{champion2003obd,savary2012obd,zhitomirsky2012obd}, where
the leading energetic effects~\footnote{
  Expected to be biquadratic exchange~\cite{savary2012obd} for \garnet{} and multi-spin
  interactions~\cite{savary2012obd,mcclarty2009energetic,rau2016vcff} for \eto{}.
} are
na\"ively~\cite{mcclarty2009energetic,petit2014gap,rau2016vcff} expected to be
small~\cite{savary2012obd}.  In this context, the pseudo-Goldstone gap provides
a quantitative benchmark which may be used to distinguish order-by-quantum
disorder from more conventional energetic selection.
Given knowledge of the models for these
materials~\cite{brueckel1988dynamical,savary2012obd} and when
semi-classical picture is a good description, our result provides a
straightforward way to estimate the pseudo-Goldstone gap observed
experimentally, cleanly demonstrating the utility of these results.
 
\begin{figure}[tp]
  \includegraphics[width=\columnwidth]{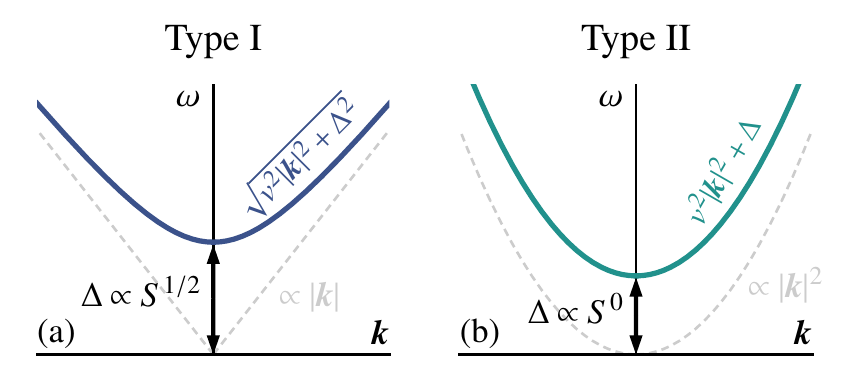}
  \caption{\label{fig:types}
    Schematic form of the spectrum of (a) type I and (b) type II
pseudo-Goldstone modes.  Type I modes have $\omega \sim |\vec{k}|$ at
$O(S)$, with the pseudo-Goldstone gap scaling as $\Delta \sim
O(S^{1/2})$, while type II modes have $\omega \sim |\vec{k}|^2$ at
$O(S)$, with $\Delta \sim O(S^0)$.
  }
\end{figure}
\sectitle{Spin-wave theory} We first review the physics of
pseudo-Goldstone modes as they appear in linear spin-wave theory. As for
the usual Goldstone modes, these can be classified into two
types~\cite{brauner2010spontaneous,wantanabe2012goldstone}, which we
denote as I and II, which correspond to having
non-conserved and conserved order parameters, respectively. For a type I
pseudo-Goldstone mode the linear spin-wave dispersion vanishes
linearly $\sim |\vec{k}|$, while for the type II case it vanishes
quadratically $\sim |\vec{k}|^2$, as illustrated in
Fig.~\ref{fig:types}.

More explicitly, we can define
the linear spin-wave Hamiltonian~\cite{holstein1940lswt}
\begin{equation}
  S \sum_{\vec{k}} \sum_{\alpha\beta} \left[ A^{\alpha\beta}_{\vec{k}} \h{a}_{\vec{k}\alpha} a^{}_{\vec{k}\beta}+
    \frac{1}{2}\left(B^{\alpha\beta}_{\vec{k}} \h{a}_{\vec{k}\alpha} \h{a}_{-\vec{k}\beta}+\hc\right)
    \right],
\end{equation}
where $a_{\vec{k}\alpha}$ is the (Fourier-transformed)
Holstein-Primakoff boson with wave-vector $\vec{k}$ on sublattice
$\alpha$ of the (magnetic) unit cell.  The matrices
$\mat{A}_{\vec{k}}$ and $\mat{B}_{\vec{k}}$ depend on the classical
ordering pattern and the exchange model; see \ac{SM}~\cite{supp} for
details. The linear spin-wave spectrum is determined by the
eigenvalues of the Bogoliubov dispersion
matrix~\cite{blaizot1986quantum}
\begin{equation}
  \mat{\sigma}_3 \mat{M}_{\vec{k}} \equiv
  \left(
    \begin{array}{cc}
      \mat{A}_{\vec{k}} & \mat{B}_{\vec{k}} \\
      -\h{\mat{B}}_{\vec{k}} & -\trp{\mat{A}}_{-\vec{k}} \\
    \end{array}
  \right),
\end{equation}
where $\mat{\sigma}_3 \equiv {\rm diag}(+\mat{1},-\mat{1})$ is a block
Pauli matrix.  A pseudo-Goldstone mode appears as a zero in the linear
spin-wave spectrum. Without loss of generality, we assume that
this zero mode lies at the zone center, with $\mat{M}_{0}$ being
positive semi-definite, and $\mat{M}_{\vec{k}}$ positive definite
elsewhere (this is always possible for commensurate magnetic orders).

\begin{figure}[tp]
  \centering
  \includegraphics[width=0.9\columnwidth]{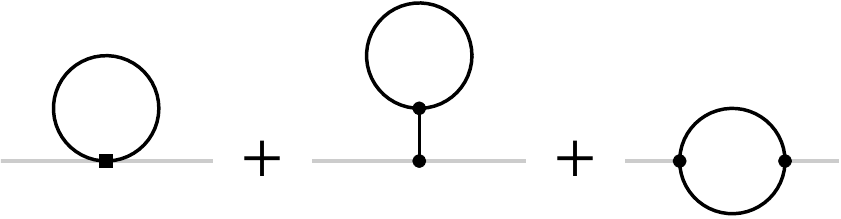}
  \caption{The three classes of (Hugenholtz) diagrams contributing
to the  $O(S^0)$ magnon self-energy,
$\mat{\Sigma}^R(\vec{k},\omega)$.  The (free) propagators
($\textbf{---}$) include normal and anomalous parts, and are connected
to external legs ($\textcolor{cG}{\textbf{---}}$). The first diagram
involves a single four-magnon interaction ({\scriptsize
$\blacksquare$}), while the second and third involve a pair of
three-magnon interactions ($\bullet$).
    \label{fig:diagrams}
  }
\end{figure}
When spin-wave interactions are included, the excitation energies are
indicated by the poles of the (retarded) magnon Green's function
\begin{equation}
  \label{eq:green}
  \mat{G}^R(\vec{k},\omega) \equiv \left[
    (\omega + i0^+)\mat{\sigma}_3
    -S\mat{M}_{\vec{k}}  -\mat{\Sigma}^R(\vec{k},\omega)
    \right]^{-1},
\end{equation}
where $\mat{\Sigma}^R(\vec{k},\omega)$ is the (retarded) self-energy.
We use a formalism where the free magnon Green's function is defined
as a matrix that includes both the sublattice indices and the normal
and anomalous contributions~\cite{blaizot1986quantum}.  The effects of
spin-wave interactions encoded in the self-energy can be computed
perturbatively~\cite{negele1988quantum} in the limit $1/S \rightarrow
0$~\cite{dyson1956rmp,harris1971}. The leading contributions at order
$O(S^0)$ are illustrated in Fig.~\ref{fig:diagrams}

The determination of the poles of the magnon Green's function then
proceeds perturbatively in the self-energy, with respect to
$\vec{M}_{\vec{k}}$.  For the type I case, one finds the relevant
low-energy subspace of $\mat{\sigma}_3\mat{M}_0$ is similar to a
defective Jordan block~\cite{blaizot1986quantum}. The pseudo-Goldstone
gap, $\Delta$, at leading order is then
\begin{equation}
  \label{eq:pg-one}
  \Delta =S^{1/2}
  \sqrt{
    2 \re \left[
      \h{\mat{V}}_0\mat{\Sigma}^R(\vec{0},0) \mat{\sigma}_3 \mat{M}_0 \vec{V}_0
    \right]
  } + O(S^{-1/2}),
\end{equation}
where $\vec{V}_0$ characterizes part of the zero mode
subspace~\cite{supp}. For the type II case, there are two linearly
independent eigenvectors of $\mat{\sigma}_3\mat{M}_0$ with eigenvalue
zero, $\vec{V}_0$ and $\vec{W}_0$.  The pseudo-Goldstone gap is then
\begin{equation}
  \label{eq:pg-two}
\Delta = 
\sqrt{
      \left(\h{\mat{V}}_0\mat{\Sigma}^R(\vec{0},0)\vec{V}^{}_0\right)^2 -
      \left|\h{\mat{V}}_0\mat{\Sigma}^R(\vec{0},0)\vec{W}^{}_0\right|^2
      } + O(S^{-1}),
\end{equation}
at leading order. Once the self-energy due to magnon-magnon
interactions is computed, these equations
[Eqs.~(\ref{eq:pg-one},\ref{eq:pg-two})] allow the direct calculation
of the pseudo-Goldstone gap.  A detailed derivation for both cases is
provided in the \ac{SM}~\cite{supp}.

These follow distinct scalings with the spin length [see
Eqs.(\ref{eq:pg-one},\ref{eq:pg-two})]: for a type I pseudo-Goldstone
mode the gap scales as $\Delta \sim O(S^{1/2})$, while for a type II
pseudo-Goldstone mode it scales as $\Delta \sim O(S^0)$. Away from the
zone center~\cite{wantanabe2012goldstone}, the spectrum takes the
low-energy form $\sim \sqrt{v^2|\vec{k}|^2 + \Delta^2}$ for type I
modes, while for type II modes it takes the form $\sim v^2|\vec{k}|^2
+ \Delta$, as shown schematically in Fig.~\ref{fig:types}.

\begin{table*}[tp]
  \begin{tabular}{lccccccc}
    \toprule
    Model / Material & Parameters & Type & $\Delta$
    & $\left(\frac{\del^2 \epsilon}{\del \theta^2}\right)_0$ & $\left(\frac{\del^2 \epsilon}{\del \phi^2}\right)_0$  & $S^{-1}\sqrt{\left(\frac{\del^2 \epsilon}{\del \theta^2}\right)_0\left(\frac{\del^2 \epsilon}{\del \phi^2}\right)_0}$
    & $S=\frac{1}{2}$ / Exp.\\
    \midrule    
    \addlinespace
    Heisenberg-compass &$|K|/|J| \ll 1$& I  & $0.52 S^{\frac{1}{2}} |K|^{\frac{3}{2}}/|J|^{\frac{1}{2}} $ & $2|K|S^2$& $0.137  K^2 S/|J| $& $0.52  S^{\frac{1}{2}}|K|^{\frac{3}{2}}/|J|^{\frac{1}{2}}$ \\
    (Square, Ferro.) &$K/|J| = -0.5$ &I & $0.17 |J| S^{\frac{1}{2}}$& $|J| S^2$& $0.0286 |J| S$& $0.17 |J|S^{\frac{1}{2}}$ \\
    \addlinespace
    Heisenberg-compass~\cite{belorizky1980calculation} &$|K|/|J| \ll 1$ & II & $0.093 K^2/|J|$ & $0.093 K^2S/|J|$ & $0.093 K^2 S/|J|$ & $0.093 K^2/|J|$\\
    (Cubic, Ferro.) &$K/|J| = +0.5 $  & II &$0.030|J|$ & $0.030|J|S$& $0.030|J|S$ & $0.030|J|$\\
    &$K/|J| = -0.5 $ & II & $0.024|J|$  &$0.024|J|S$ & $0.024|J|S$ & $0.024|J|$ \\
    \addlinespace
    \addlinespace
    Heisenberg-Kitaev~\cite{chaloupka2010kh} & $|K| \ll |J|$ & II & $0.0897 K^2/|J|$&  $0.0897 K^2 S/|J|$ &  $0.0897 K^2 S/|J|$   &  $0.0897 K^2/|J|$\\
    (Honeycomb,  Ferro.)
         & $K/|J| = -2.0$ & II & $0.208 |J|$ & $0.208 |J|S$ & $0.208 |J|S$ & $0.208 |J|$\\
          & $K/|J| = -0.65$ & II & $0.03 |J|$ & $0.0300|J|S$ & $0.0300|J|S$ & $0.0300|J|$ & $\sim 0.05|J|$~\cite{gohlke2017dynamics}\\            
    \addlinespace
    Heisenberg-Kitaev~\cite{chaloupka2010kh} & $|K| \ll J$ & I+I
                              & $0.83 |K|S^{\frac{1}{2}}$ & $2(3J + K)S^2$ & $0.115 K^2S/J$ & $0.83 |K|S^{\frac{1}{2}}$\\
    (Honeycomb, N\'eel) & $K/J = +2.0 $ & I+I
                              & $1.66 J S^{\frac{1}{2}}$ & $10 JS^2$ & $0.274 J S$  & $1.66 J S^{\frac{1}{2}}$\\
          & $K/J = -0.5$ & I+I & $0.434 J S^{\frac{1}{2}}$ & $5JS^2$ & $0.038 JS$  &$0.434 J S^{\frac{1}{2}}$\\
    \addlinespace
    Heisenberg-$\Gamma$~\cite{rau2014jkg}
          & $\Gamma \ll |J|$ & I
                              & $0.29\Gamma^{2}/|J|^{} S^{\frac{1}{2}}$&  $3\Gamma S^2$ &  $0.028\Gamma^3 S/|J|^2$   &  $0.29\Gamma^{2}/|J| S^{\frac{1}{2}}$\\
    (Honeycomb, Ferro.)   & $\Gamma/|J| = +0.5 $ & I
                              & $0.081|J|S^{\frac{1}{2}}$ & $1.5|J|S^2$ & $0.00437|J|S$ &  $0.081|J|S^{\frac{1}{2}}$\\
          & $\Gamma/|J| = +1.0$ & I
                              & $0.355|J|S^{\frac{1}{2}}$ & $3|J|S^2$ & $0.042|J|S$ & $0.355|J|S^{\frac{1}{2}}$\\        
    \addlinespace
    \addlinespace
    $J_1$-$J_2$~\cite{henley1989,chandradoucot1988,chandracoleman1990}
          & $J_1/J_2 \ll 1$ & I+I & {$1.44 J_1 S^{\frac{1}{2}}$} & {$4(2J_2-J_1)S^2$} & $0.2604 J_1^2 S/J_2$ & $1.44 J_1 S^{\frac{1}{2}}$ & \\
    (Square, Stripe)
          & $J_1/J_2 = 0.5$ & I+I & {$0.63 J_2 S^{\frac{1}{2}}$}& {$6J_2 S^2$} &$0.0668 J_2S$ & $0.63 J_2 S^{\frac{1}{2}}$ & $0.61J_2S^{\frac{1}{2}}$~\cite{singh2003gap}\\
          & $J_1/J_2 = 1$   & I+I & {$1.08 J_2 S^{\frac{1}{2}}$} & {$4J_2 S^2$} &$0.294 J_2S$ & $1.08J_2S^{\frac{1}{2}}$ & $0.96J_2S^{\frac{1}{2}}$~\cite{singh2003gap}\\
    \addlinespace
    $J_1$-$J_2$~\cite{chandradoucot1988,chubukov1992triangular}
          & $J_2/J_1 = 0.25$ & II+II
                              & {$0.53J_1$} & {$0.53J_1S$} & $0.53J_1S$ & $0.53J_1$ & \\
    (Triangular, Stripe)
          & $J_2/J_1 = 0.5$ & II+II
                              & {$0.45J_1$}& {$0.45J_1S$} &$0.45J_1S$ & $0.45J_1$ &  \\
          & $J_2/J_1 = 0.75$ & II+II
                              & {$0.58J_1$}& {$0.58J_1S$} &$0.58J_1S$ & $0.58J_1$ &  \\
    \addlinespace
    \midrule
    \addlinespace
    \eto{}~\cite{champion2003obd,savary2012obd,zhitomirsky2012obd}
          & \citet{savary2012obd}
                                & I & $31.1 \mueV$ & $157.5 \mueV$ & $1.536 \mueV$& $31.1 \mueV$ &$43$-$53\mueV$~\cite{lhotel2017gap,ross2014gap}\\
    \addlinespace
    \garnet{}~\cite{shender1982antiferromagnetic,brueckel1988dynamical}
          & \citet{brueckel1988dynamical}
                       & I+I & $ 262 \mueV$ & $4\meV$ & $107.5\mueV$ & $262 \mueV$ & $136 \mueV$~\cite{brueckel1988dynamical}\\
    \addlinespace
    \bottomrule
  \end{tabular}
  \caption{
    \label{tab:results}
    Calculations showing the equality of the pseudo-Goldstone gap,
$\Delta$ computed from non-linear spin-wave theory
[Eqs.~(\ref{eq:pg-one},\ref{eq:pg-two})] and then independently from
the curvatures of the classical and quantum zero-point energies
[Eq.~(\ref{eq:curvature-formula})]. For each model, the lattice, the
exchange regime, the type of pseudo-Goldstone mode and several choice
of parameters are listed.  When available, additional theoretical or
experimental estimates of the pseudo-Goldstone gap are shown.
  }
\end{table*}

\sectitle{Curvature formula and semi-classical dynamics}
We now motivate the curvature formula for the pseudo-Goldstone gap,
$\Delta$ [Eq.~(\ref{eq:curvature-formula})], through a heuristic
semi-classical argument. It is useful to construct a
local frame $(\vhat{x}_{\alpha},\vhat{y}_{\alpha},\vhat{z}_{\alpha})$ where
$\vhat{z}_{\alpha}$ is the ordering direction, $\vhat{x}_{\alpha}$ is
the soft-mode direction and $\vhat{y}_{\alpha} = \vhat{z}_{\alpha}
\times \vhat{x}_{\alpha}$
If we parametrize the
soft mode by an angle $\phi$, and the (locally) orthogonal directions
by an angle $\theta$, we can define the 
classical spin configuration
\begin{equation}
  \label{eq:soft-mode}
  \vec{S}_{\alpha}  = S\left(\phi\vhat{x}_{\alpha} + \theta \vhat{y}_\alpha
  +\vhat{z}_{\alpha} \left[1-(\phi^2+\theta^2)\right]^{1/2}\right),
\end{equation}
accurate to quadratic order in $\theta$ and $\phi$. For simplicity, we
have assumed here that the soft mode is uniform, with the relative
weight of the rotations not varying between sublattices -- this
assumption is not essential, and can be lifted~\cite{supp}.
These variables have the Poisson bracket $\{\phi,\theta\}
= NS$ and thus essentially behave like a position and its canonically
conjugate momentum. For the type I case, $\phi$ is classically soft,
with no restoring force, while is $\theta$ is not, while for the type II
case both are classically soft.

If we treat these collective coordinates as
classical dynamical variables, then quantum fluctuations can be included in an ad-hoc way
by using the semi-classical spin-wave energy density $\epsilon(\theta,\phi)$ as an
effective potential; explicitly,
\begin{equation}
  \label{eq:zpm}
  \epsilon(\phi,\theta) \equiv \begin{cases}
    S^2\epsilon_{\rm cl}(\theta) + S\epsilon_{\rm qu}(\phi,0), & {\rm type\ I}\\
    S\epsilon_{\rm qu}(\phi,\theta), & {\rm type\ II}
  \end{cases},
\end{equation}
where $\epsilon_{\rm cl}$ is the classical energy density and
$\epsilon_{\rm qu}$ is the quantum zero-point energy
density~\cite{supp} computed in linear spin wave theory for the soft
spin configurations [Eq.~(\ref{eq:soft-mode})].  Since the $\theta$
direction is not soft for type I pseudo-Goldstone modes, the classical
part of the energy must be included.  The quantum zero-point energy
density is defined as
$
  \epsilon_{\rm qu}(\phi,\theta)~\equiv~(2 N)^{-1}\sum_{\vec{k}\alpha}\epsilon_{\vec{k}\alpha}(\phi,\theta),
$
where $\epsilon_{\vec{k}\alpha}(\phi,\theta)$ are the spin-wave
energies found expanding about a classical ground state with finite
$\phi,\theta$.  For the type II case, this zero-point energy is
well-defined for arbitrary $\phi$ and $\theta$, with the classical
energy independent of both variables, while for the type I case, the
zero-point energy is ill-defined for $\theta \neq 0$, and thus we 
fix $\theta=0$.

The curvatures of the total semi-classical energy density
directly determine the normal mode frequency of
$\theta$ and $\phi$ via the classical equations of motion~\cite{smitbelgers}
\begin{subalign}
  \frac{d\phi}{dt} &= +\frac{1}{S}\frac{\del \epsilon}{\del \theta} \approx  +\left(\frac{\del^2 \epsilon}{\del \theta \del \phi}\right)_0 \phi + \left(\frac{\del^2 \epsilon}{\del \theta^2}\right)_0 \theta,   \\
  \frac{d\theta}{dt} &= -\frac{1}{S}\frac{\del \epsilon}{\del \phi} \approx  -\left(\frac{\del^2 \epsilon}{\del \phi^2}\right)_0 \phi -\left(\frac{\del^2 \epsilon}{\del \theta \del \phi}\right)_0 \theta, 
\end{subalign}
giving the pseudo-Goldstone gap shown in Eq.~(\ref{eq:curvature-formula}).
For all the cases of interest the cross term vanishes, so we omit
$\left(\frac{\del^2 \epsilon}{\del \theta \del \phi}\right)_0$ in what follows.   Multiple sets of
pseudo-Goldstone modes can be handled in a similar fashion, reducing
to multiple independent copies of either the type I or type II
structures described above (absent fine-tuning). 

We have computed the pseudo-Goldstone gap for a wide variety of
models~\cite{supp} using both non-linear spin-wave theory 
[Eqs.~(\ref{eq:pg-one},\ref{eq:pg-two})], and using the curvature
formula [Eq.~(\ref{eq:curvature-formula})], which involves
only linear spin-wave theory. The results are presented in
Table~\ref{tab:results}, where one can see that the two methods agree
\emph{exactly} for all models considered. This includes two- and
three-dimensional models, isotropic and anisotropic models, models
with and without magnon decay as well as realistic models for two
experimental order-by-quantum-disorder candidates. Some details for
each model, as well as examples of how to define $\phi$, $\theta$ and
compute the curvatures are provided in the \ac{SM}~\cite{supp}.

\sectitle{Proof of formula} The equivalence between these two
approaches can be understood as follows. The essential ingredient is
to notice that the Holstein-Primakoff expansion should not depend on
the choice of initial classical ground state about which one expands,
so long as it is sufficiently close to the true ground state of the
model. In other words, the expansion must ``self-correct'', with the
expectation values of the magnons giving the appropriate true ground
state spin directions, order-by-order in $1/S$.
For a model without any accidental degeneracies this can be shown at
$O(S)$~\cite{supp}; the required cancellation relates the magnon
energy at zero wave-vector to the curvature of the classical energy
density, as in the formula of \citet{smitbelgers}.

To understand the implication of this self-correction at higher order,
we must proceed more indirectly.  Define the rotated Hamiltonian
$
\mathcal{H}(\phi,\theta) = \h{U(\phi,\theta)} H U(\phi,\theta),  
$
where $U(\phi,\theta)$ produces the soft configurations
of Eq.~(\ref{eq:soft-mode}) from the state defined by
$\vhat{z}_{\alpha}$. The self-correction condition is then the
(trivial) fact that the ground state energy of
$\mathcal{H}(\phi,\theta)$ is independent of $\phi$ and $\theta$. If
one considers the implications of this statement on the
\emph{derivatives} of the ground state energy of
$\mathcal{H}(\phi,\theta)$, at second order one finds that this
implies the sum rule~\cite{viswanath1994recursion,supp}
\begin{equation}
  \label{eq:sumrule}
  \h{\vec{U}}_\mu \mat{\sigma}_3\left[
    \int d\omega\ \omega\ \mat{A}(\vec{0},\omega)\right] 
  \mat{\sigma}_3\k{\vec{U}}_\nu =\frac{1}{SN}
  \left\langle
   \left( \frac{\del^2 \mathcal{H}}{\del \lambda_\mu \del \lambda_\nu}\right)_0
    \right\rangle  ,
\end{equation}
where $\mu,\nu = \Theta,\Phi$, $\lambda_\Phi = \phi$ and
$\lambda_\Theta = \theta$ and we define $\vec{U}_{\Phi} \equiv
(\vec{V}_0-\vec{W}_0)/(i\sqrt{2})$, $\vec{U}_{\Theta} = (\vec{V}_0
+\vec{W}_0)/\sqrt{2}$ which span the zero-mode subspace~\cite{supp}. The
magnon spectral function is defined as 
$
  \mat{A}(\vec{k},\omega)~\equiv~(2i)^{-1}\left[\mat{G}^R(\vec{k},\omega) - \mat{G}^A(\vec{k},\omega)\right],
$
where the $\vec{G}^A(\vec{k},\omega) \equiv
\h{\mat{G}^R(\vec{k},\omega)}$ is the advanced magnon Green's
function.

Using this sum rule [Eq.~(\ref{eq:sumrule})] one can show that, at
$O(S^0)$, the left-hand side is directly related to the
pseudo-Goldstone gap, while the right-hand side is related to the
curvatures of the classical energy density and quantum zero-point
energy density at $O(S^2)$ and $O(S)$, respectively~\cite{supp}.  This
argument does not directly extend to higher orders in $1/S$ or to
computing the energies of finite energy modes to $O(S^0)$. We note
that in broad strokes this argument bears some resemblance to the
Witten-Veneziano formula~\cite{witten1979current,veneziano1979u} for
the mass of the $\eta'$-meson. In addition, the sum rule
[Eq.~(\ref{eq:sumrule})] is related to Dashen's formula
~\cite{dashen1} for the mass of pseudo-Goldstone bosons, such as the
pion, when chiral symmetry is broken~\cite{supp}.

\sectitle{Discussion} We now discuss some applications to two leading experimental
candidates for order-by-quantum-disorder. The first of these is the compound
\garnet{}, which is a three-dimensional $S=5/2$ version of
one of the canonical order-by-disorder
models, the $J_1$-$J_2$ model~\cite{chandradoucot1988,henley1989}.
This system has a pair of type I pseudo-Goldstone
modes, as well as two true Goldstone modes. Due to the low-symmetry of
the lattice, there are several independent (isotropic) couplings,
which have been estimated by comparison of the predictions of
linear spin-wave theory with the inelastic
neutron scattering spectrum at zero field~\cite{brueckel1988dynamical}. One finds
that the gap predicted by non-linear spin-wave theory, $\sim 262
\mueV$, is of the right order of magnitude, but larger than the $136\mueV$~\cite{brueckel1988dynamical} observed
experimentally~\footnote{
Our result mostly agrees with previous
estimates~\cite{shender1982antiferromagnetic,brueckel1988dynamical} of
this gap. We note that the methods used in
Ref.~[\onlinecite{shender1982antiferromagnetic}] are non-standard, and
the stated final numerical result was in error, and revised in
subsequent work~\cite{brueckel1988dynamical}. If we restrict ourselves
to $J_1 = J_1'$ (see \ac{SM}~\cite{supp}), as was done in
Ref.~[\onlinecite{shender1982antiferromagnetic}], then our results
agree with those stated in Ref.~[\onlinecite{brueckel1988dynamical}],
once the double counting of the bonds has been accounted for.
Including $J_1' \neq J_1$ gives some quantitative disagreement with
Ref.~[\onlinecite{brueckel1988dynamical}], but they remain in
qualitative agreement.
}. This demonstrates sharply the utility of this method.
with the straightforward curvature calculation lending credence to our
more involved non-linear spin-wave result. We also note that the large
size of the predicted gap supports the picture that \garnet{} is truly
an example of order-by-quantum-disorder, and perhaps energetic
corrections, such as biquadratic interactions~\cite{savary2012obd},
are small. This quantitative disagreement could be due to several
factors, such as the need to include additional anisotropic or
biquadratic exchanges in the model or the need to include interaction
or thermal effects in fitting the exchange parameters, which were done
at zero magnetic field and moderate temperature~\cite{supp}

Finally we turn to \eto{}, one of the more ideal material platforms
for finding order-by-quantum-disorder~\cite{savary2012obd,zhitomirsky2012obd}. This
is a three-dimensional $S=1/2$ XY antiferromagnet with a single type I
pseudo-Goldstone mode. Using the exchange parameters of
Ref.~[\onlinecite{savary2012obd}], we find that the gap, computed
directly in non-linear spin-wave theory as well as via the curvature
formula is $\sim 31\mueV$~\footnote{Our result differs from the estimate presented in
Ref.~[\onlinecite{savary2012obd}] by a factor of $\sqrt{2}$. The simplicity of the
curvature estimate presented here, relative to that carried out in
Ref.~[\onlinecite{savary2012obd}], as well as the agreement with our
explicit non-linear spin-wave calculation, leads us to trust our value of
$31 \mueV$ over the value $22 \mueV$ reported in
Ref.~[\onlinecite{savary2012obd}].}. This theoretical value is still
somewhat lower than the $43\mueV$~\cite{petit2014gap,lhotel2017gap} and $53\mueV$~\cite{ross2014gap} that have been reported
experimentally in \eto{}. This disagreement could be the result of
several factors, such as spin-wave theory being non-quantitative~\cite{maryasin2014obd} at
$S=1/2$, uncertainties in the exchange parameters~\cite{savary2012obd} or the presence of energetic
corrections~\cite{mcclarty2009energetic,rau2016vcff} in addition to the 
order-by-quantum-disorder contribution.

There are many other experimental systems where
order-by-quantum-disorder may be lurking, and where the results
presented here would be useful. In the same vein as \eto{},
order-by-quantum-disorder may play a role in the pyrochlores
\abo{Yb}{Ti} and its cousin
\abo{Yb}{Ge}~\cite{jaubert2015multiphase,petit2015dynamics}, and
perhaps even in the ytterbium based
spinels~\cite{lau2005,higo2017spinels1,ybspinel2,rau2018frustration}. Order-by-disorder
has also has played a key role in the understanding of
models~\cite{chaloupka2010kh,lee2014obd,ioannis2015} of Kitaev
materials~\cite{winter2017models} such as Na\tsub{2}IrO\tsub{3},
$\alpha$-RuCl\tsub{3} and
$(\alpha,\beta,\gamma)$-Li\tsub{2}IrO\tsub{3}, and also 
in related strongly spin-orbit coupled compounds~\cite{jackeli2015triangular,aczel2016fcc,ye2017obd}.

On the more theoretical front, one could ask whether the methods
discussed could also resolve larger degeneracies, e.g.  sub-extensive
line or surface
degeneracies~\cite{yildrim1998frustration,hizi2007effective,bergman2007order,stasiak}.
More drastically, one could consider a case like the the kagom\'e
anti-ferromagnet, where the classical ground state is macroscopically
degenerate~\cite{kagome1,kagome2,kagome3}. If one expands about states
that are expected to be selected by $1/S$ corrections, one finds that
the linear spectrum has a large number of zero
modes~\cite{harris1992zero}. It would be interesting to compare the
semi-classical approach outlined here to the approaches followed in
Ref.~[\onlinecite{chubukov1992kagome}].

\sectitle{Acknowledgments}
We thank M. Zhitomirsky for a helpful discussion, M. Gingras for
useful comments and R. Coldea for collaborations on related topics.  This
work was in part supported by Deutsche Forschungsgemeinschaft (DFG)
under grant SFB 1143.

\bibliography{draft}

\clearpage

\addtolength{\oddsidemargin}{-0.75in}
\addtolength{\evensidemargin}{-0.75in}
\addtolength{\topmargin}{-0.725in}

\newcommand{\addpage}[1] {
 \begin{figure*}
   \includegraphics[width=8.5in,page=#1]{supp.pdf}
 \end{figure*}
}
\addpage{1}
\addpage{2}
\addpage{3}
\addpage{4}
\addpage{5}
\addpage{6}
\addpage{7}
\addpage{8}
\addpage{9}
\addpage{10}
\addpage{11}
\addpage{12}
\addpage{13}
\addpage{14}
\addpage{15}
\addpage{16}
\addpage{17}
\addpage{18}
\addpage{19}
\addpage{20}
\addpage{21}
\addpage{22}
\addpage{23}
\addpage{24}
\addpage{25}
\addpage{26}
\addpage{27}
         
\end{document}